\documentclass[useAMS,usegraphicx,usenatbib,usedcolumn]{mn2e}
\usepackage{rotating}
\usepackage{lscape}
\usepackage{acronym}
\usepackage{subfigure}
\usepackage[dvips]{graphicx}
\usepackage{epsf}
\usepackage{graphics,latexsym,longtable,lscape}

\newif\ifAMStwofonts
\AMStwofontstrue

\newcommand{\HII}{H\,{\sc ii}}
\newcommand{\OIII}{[O\,{\sc iii}]}

\newcommand{\SII}{[S\,{\sc ii}]}
\newcommand{\Halpha}{H${\alpha}$}

\title[Radio PNe in the Magellanic Clouds]{Radio Planetary Nebulae in the Magellanic Clouds}
\author[Filipovi\'c et al.]{M. D. Filipovi\'c,$^1$\thanks{Email: m.filipovic@uws.edu.au}
M. Cohen,$^2$
W. A. Reid,$^3$
J. L. Payne,$^1$
Q. A.~Parker,$^3$
\newauthor
E. J. Crawford,$^1$
I. S. Boji\v{c}i\'c,$^3$
A. Y. De Horta,$^1$
A. Hughes,$^{4,5}$
J. Dickel,$^6$
and
\newauthor
F. Stootman$^1$\\
$^1$University of Western Sydney, Locked Bag 1797, Penrith South DC, NSW 1797, Australia\\
$^2$Radio Astronomy Laboratory, University of California, Berkeley, CA 94720\\
$^3$Department of Physics, Macquarie University, Sydney, NSW 2109, Australia\\
$^4$Swinburne University of Technology, Hawthorn, VIC 3122, Australia\\
$^5$CSIRO Australia Telescope National Facility, PO Box 76, Epping, NSW 1710, Australia\\
$^6$Department of Physics and Astronomy, University of New Mexico, 800 Yale Blvd NE, Albuquerque, NM 87131, USA
}
\begin{document}

\date{Accepted 2009 June XX. Received 2009 March XX; in original form 2009 March XX}

\pagerange{\pageref{firstpage}--\pageref{lastpage}} \pubyear{2009}

\maketitle

\label{firstpage}

\begin{abstract}
We report the extragalactic radio-continuum detection of 15 planetary nebulae (PNe) in the Magellanic Clouds (MCs) from recent Australia Telescope Compact Array+Parkes mosaic surveys. These detections were supplemented by new and high resolution radio, optical and IR observations which helped to resolve the true nature of the objects. Four of the PNe are located in the Small Magellanic Cloud (SMC) and 11 are located in the Large Magellanic Cloud (LMC). Based on Galactic PNe the expected radio flux densities at the distance of the LMC/SMC are up to $\sim2.5$~mJy and $\sim$2.0~mJy at 1.4~GHz, respectively. We find that one of our new radio PNe in the SMC has a flux density of 5.1~mJy at 1.4~GHz, several times higher than expected. We suggest that the most luminous radio PN in the SMC (N\,S68) may represent the upper limit to radio peak {luminosity} because it is $\sim$3 times more luminous than NGC\,7027, the most luminous known Galactic PN. We note that the optical diameters of these 15 MCs PNe vary from very small ($\sim0.08$~pc or 0.32\arcsec; SMP~L47) to very large ($\sim1$~pc or 4\arcsec; SMP~L83). Their flux densities peak at different frequencies, suggesting that they may be in different stages of evolution. We briefly discuss mechanisms that may explain their unusually high radio-continuum flux densities. We argue that these detections may help solve the ``missing mass problem'' in PNe whose central stars were originally 1-8 M$_\odot$. We explore the possible link between ionised halos ejected by the central stars in their late evolution and extended radio emission. Because of their higher than expected flux densities we tentatively call this PNe (sub)sample -- ``Super PNe''.
\end{abstract}


\section{Introduction}

 \label{intro}

Planetary nebulae (PNe) possess ionized, neutral, atomic, molecular and solid states of matter in diverse regions with different temperature, density and morphological structure. Their physical environments range in temperature from $10^2$~K to greater than $10^6$~K. Although these objects radiate from the \mbox{X-ray} to the radio, detected structures are influenced by selection effects due to intervening dust and gas, instrument sensitivity and distance {(see the more specific discussions by \citet{2005JKAS...38..271K,1998ApJ...499L..83D}).}

{Radio-continuum surveys of PNe in the Magellanic Clouds (MCs) potentially offer a flux-limited sample that could provide absolute physical attributes such as fluxes, emission measures, and spectral energy distributions (SEDs) (e.g., \citet{2008ApJ...681.1296Z}). These in turn are relevant to the major issues of PN evolution, although MC PNe provide very limited information on morphology.}

Most known PNe are weak thermal radio sources and although morphologies of these radio objects are similar to their optical counterparts, radio interferometric observations allow us to image the structure of a PNe's ionized component. A spherically symmetric uniform density PNe has an ionized mass, $M_{i}$, that can be expressed as:

\begin{equation}
M_{i}= 282\,(D_{kpc})^{2}{F}_{5}(n_{e})^{-1}M_{\odot},
\end{equation}

\noindent where $D_{kpc}$ is distance (kpc), $F_{5}$ is radio flux density at 5~GHz (Jy) and $n_{e}$ represents electron density (cm$^{-3}$) derived from forbidden-line ratios \citep{2000oepn.book.....K}. In cases where the PNe distance is unknown, this equation can be inverted to provide a crude but useful distance estimate \citep[][Appendix~A, eq. (A.14)]{1967ApJ...147..471M}. For example, adopting a 5~GHz flux density of 25~mJy and $T_{e} = 10^{4}$~K, gives an ionized gas mass of $0.0022 \times D^{2.5}_{kpc}M_{\odot}$. Assuming a mean PNe ionized mass of 0.1 to 0.25~$M_{\odot}$ allowed \citet{2005ApJ...627..446C} to estimate that the distance to G313.3+00.3 lies between 4.6 and 6.6~kpc. However, {this distance range may not be always valid as} Eq.~1 may not be always applicable. For example, some PNe such as NGC~7027, the brightest PN in the radio sky, exhibit an ionized mass of 0.057M$_{\odot}$ \citep{1996A&A...315L.253B}. At the other extreme, large PNe have masses up to 0.5M$_{\odot}$ or sometimes well above\footnote{Masses up to 4M$_{\odot}$ are achievable due to the interstellar medium (ISM) sweep-up \citep{2007MNRAS.382.1233W} the outer shell comes to a stand-still when about 10 times the ejected mass has been swept up.}. Therefore, we point that Eq.~1 is valid only if PN is optically thin, but this will be true in the large majority of cases.

Study of extragalactic PNe have the advantage that their distance is known with much greater certainty than those of Galactic PNe. Centimeter radio emission from PNe also can be used to estimate interstellar extinction by comparing radio and optical Balmer-line fluxes \citep*{2005ApJS..159..282L}. The study of radio PNe at a known distance allows us to better understand the properties of PNe in our own Galaxy, and ultimately to refine methods of estimating their distances. We also point out that PNe may even reflect conditions inherent in their host galaxies. However, there is consistency in the bright end cut-off of the PN Luminosity Function (PNLF) regardless of galaxy type \citep{2008ApJ...683..630H}.

We present the first complete sample (S$_{i}>1.5$~mJy in the radio-continuum of confirmed extragalactic PNe. Further analysis depends heavily on a variety of new, high resolution and time consuming observations which are underway.

\section{Radio Data}
\label{DATA}

The large majority of known Galactic PNe are weak but detectable radio-continuum objects. Their thermal radio emission is a useful tracer of nebular ionization. Because centimeter-wavelength radiation is not extinguished by dust grains, the observed emission should be a good representation of the conditions in the PNe. For these reasons, large scale radio surveys surveys of nearby galaxies such as the Magellanic Clouds (MCs) may serve as a perfect example for a detection of radio-continuum PNe outside of {our} own Milky Way.

In the past decade, several Australia Telescope Compact Array (ATCA) moderate resolution surveys of the MCs have been completed. Deep ATCA and Parkes radio-continuum \citep{1995A&AS..111..311F,1997A&AS..121..321F} and snap-shot surveys of the Small Magellanic Cloud (SMC) were conducted at 1.42, 2.37, 4.80 and 8.64~GHz by \citet{2002MNRAS.335.1085F,2005MNRAS.364..217F} and \citet{2004MNRAS.355...44P}, achieving sensitivities of 1.8, 0.4, 0.8 and 0.4~mJy~beam$^{-1}$ respectively. The maps have angular resolutions of 98\arcsec, 40\arcsec, 30\arcsec\ and 15\arcsec\ at the frequencies listed above. New complete mosaics of the SMC at 4.80 and 8.64~GHz (both at sensitivities of 0.5~mJy~beam$^{-1}$) have recently been completed by \citet{2009IAUS..256...14D} and \citet{2009IAUS..256...PDF-8}. Also, \citet{2008ApJ...688.1029M} presented a new 20-cm ATCA mosaic survey of the SMC with a resolution of 18\arcsec $\times$ 11\arcsec, which is well suited for initial PNe detection as the PNe appear as point sources.

For the Large Magellanic Cloud (LMC), a new moderate resolution (40\arcsec; sensitivity $\sim$0.6~$\mathrm{mJy~beam}^{-1}$) ATCA+Parkes survey by \citet{2006MNRAS.370..363H,2007MNRAS.382..543H} and \citet{2009SerAJ.178...65P} at 1.4~GHz ($\lambda=20$~cm) complements ATCA+Parkes mosaic images at 4.8 and 8.64~GHz obtained by \citet{2005AJ....129..790D}. For these observations, the 4.8~GHz total intensity image has a FWHM of 33\arcsec\ while the 8.64~GHz image has a FWHM of 20\arcsec. Both have sensitivities of $\sim$0.5~mJy~beam$^{-1}$ and positional uncertainties for all three radio-continuum maps of the LMC are less than 1\arcsec.

In addition to the ATCA+Parkes surveys, we also searched the 843~MHz Sydney University Molonglo Sky Survey (SUMSS; resolution $\sim$45\arcsec, sensitivity $\sim$2~mJy; \citep{1999AJ....117.1578B}) for sources co-incident with known catalogued PNe. We also searched the specific MOST (MOlonglo Synthesis Telescope) observations of the SMC presented by \citet{1998PASA...15..280T}. 

From these mosaic surveys a collection of targets was selected for follow up observation. In several sessions since 2006 we have observed 5 of these PNe with the ATCA (project C1604) in ``snap-shot'' mode at 4.8 and 8.64~GHz achieving resolutions as high as 1\arcsec. With this resolution, we expect that the larger PNe, such as SMP\,L83, might be resolved, but most PNe will still appear unresolved. These 5 PNe are marked with an $^{\star}$ in Table~1 (Column~1).

\section{Method and Results}
 \label{MaR}

\begin{figure*}
 \begin{center}
 \includegraphics[width=62mm,angle=-90]{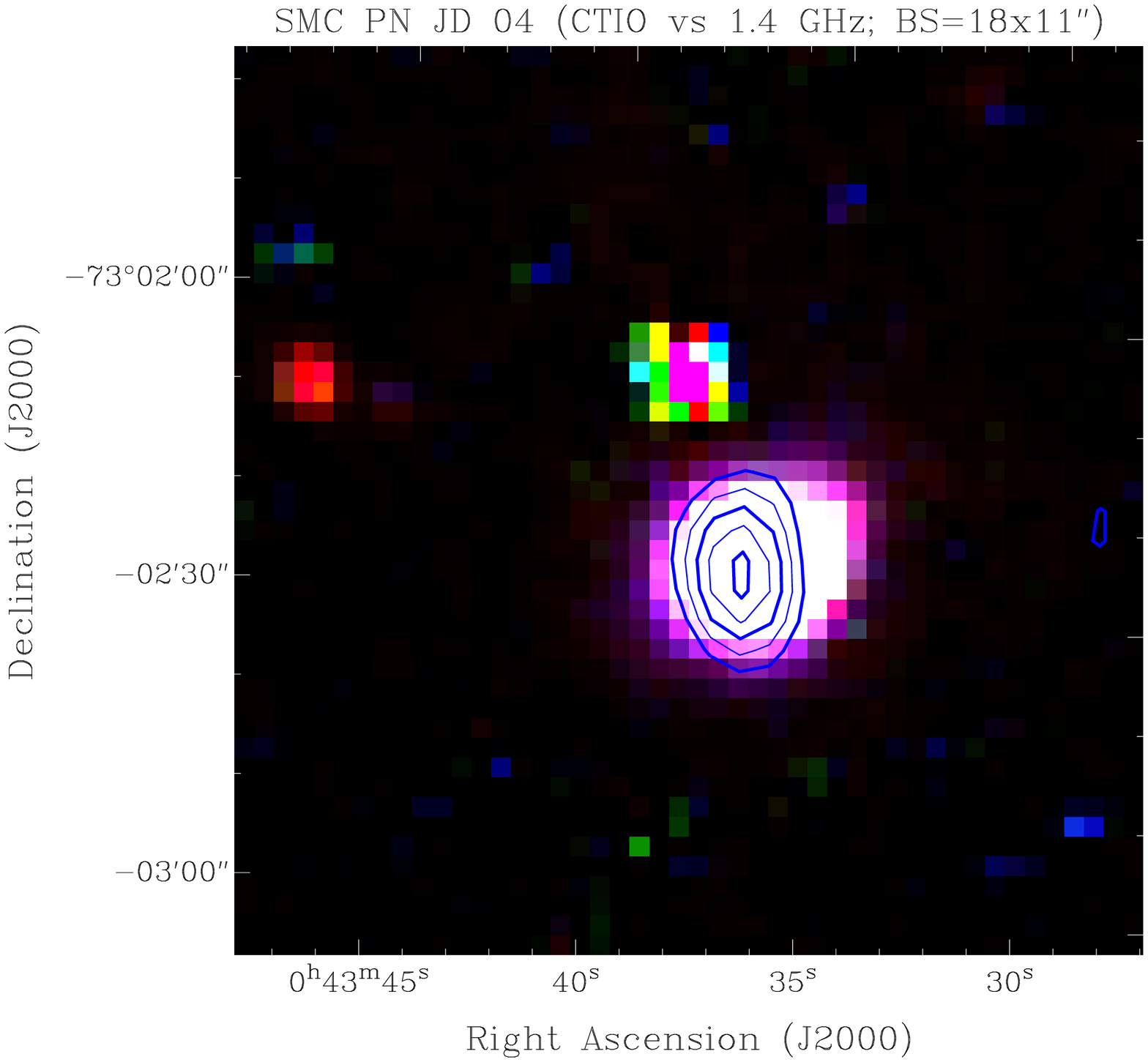}
 \includegraphics[width=62mm,angle=-90]{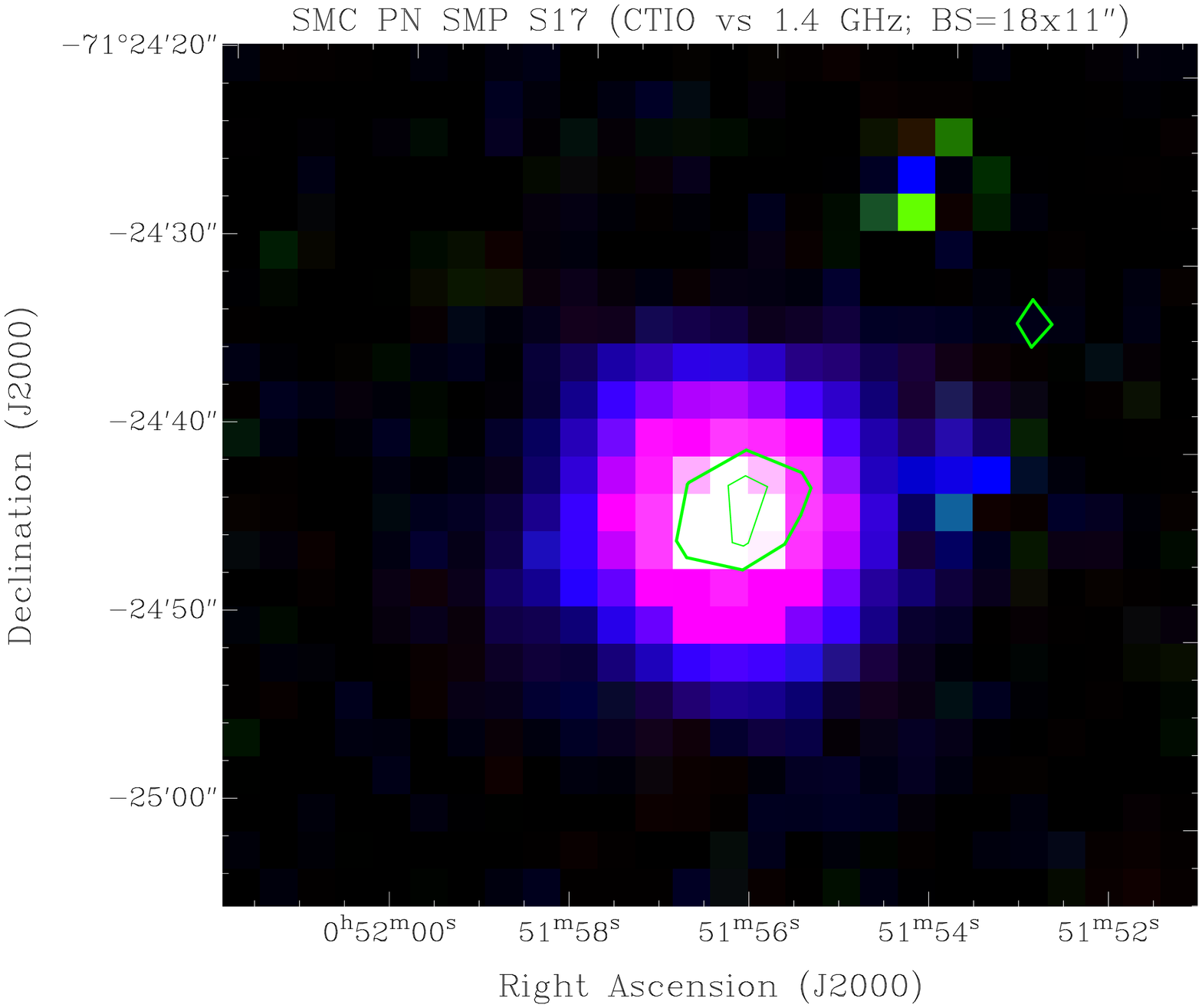}
  \caption{A composite optical image (RGB = {\Halpha}, \SII\ and \OIII, respectively,
  all continuum-subtracted) of two SMC PNe -- JD04 (left) and SMP\,S17 (right) overlaid with 1.4~GHz contours from \citet{2008ApJ...688.1029M} mosaic survey. This optical image is from the Magellanic Cloud Emission Line Survey (MCELS) (Winkler et al. in prep.). The radio-continuum contours are from 1~mJy~beam$^{-1}$ in steps of 0.5~mJy~beam$^{-1}$. The synthesised beam of 1.4~GHz survey is 18\arcsec$\times$11\arcsec. The {larger optical} extent of these two PNe is due to faint AGB halos.}
 \label{fig1-smc}
 \end{center}
\end{figure*}

\begin{figure*}
 \begin{center}
 \includegraphics[width=62mm,angle=-90]{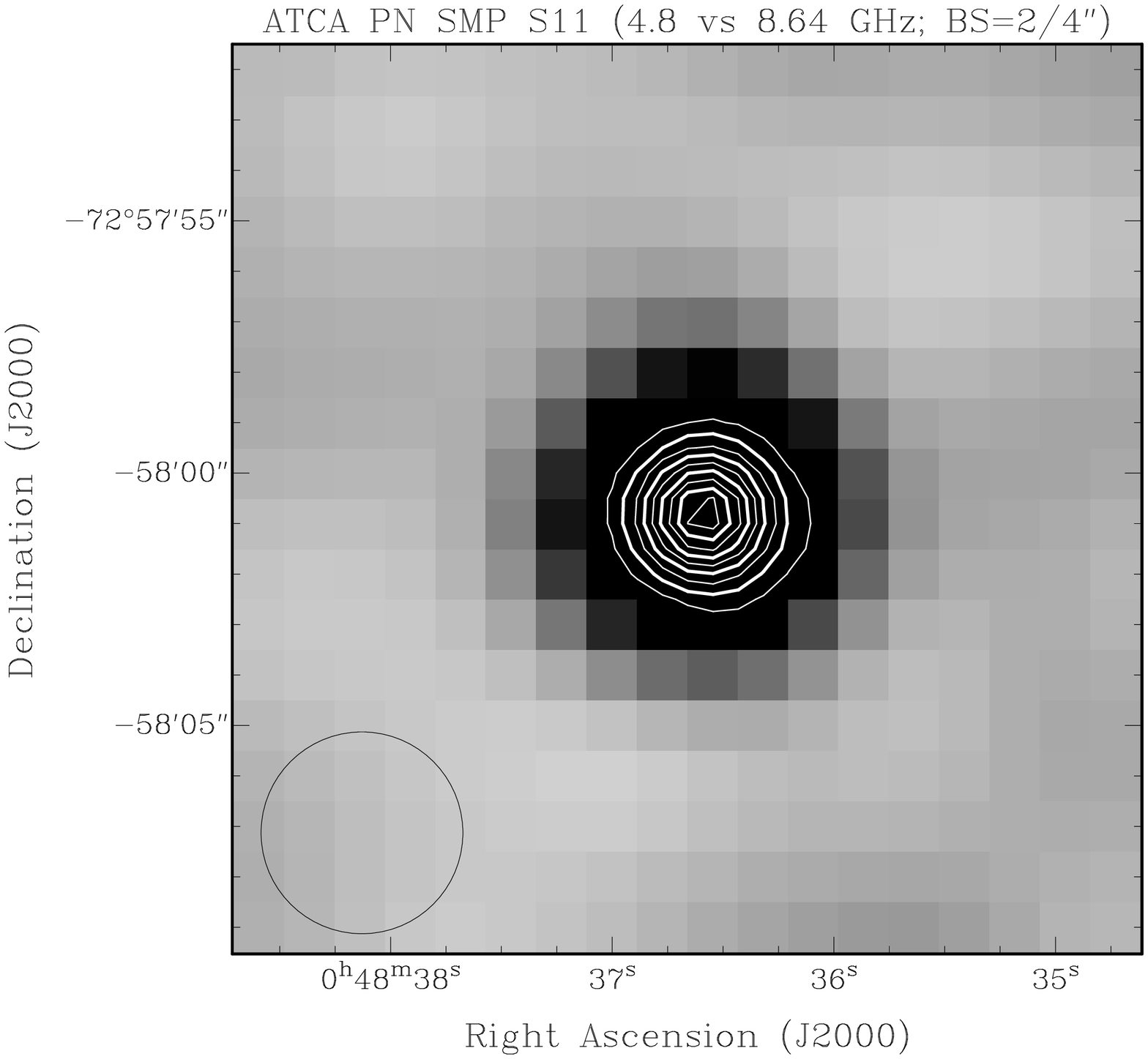}
 \includegraphics[width=62mm,angle=-90]{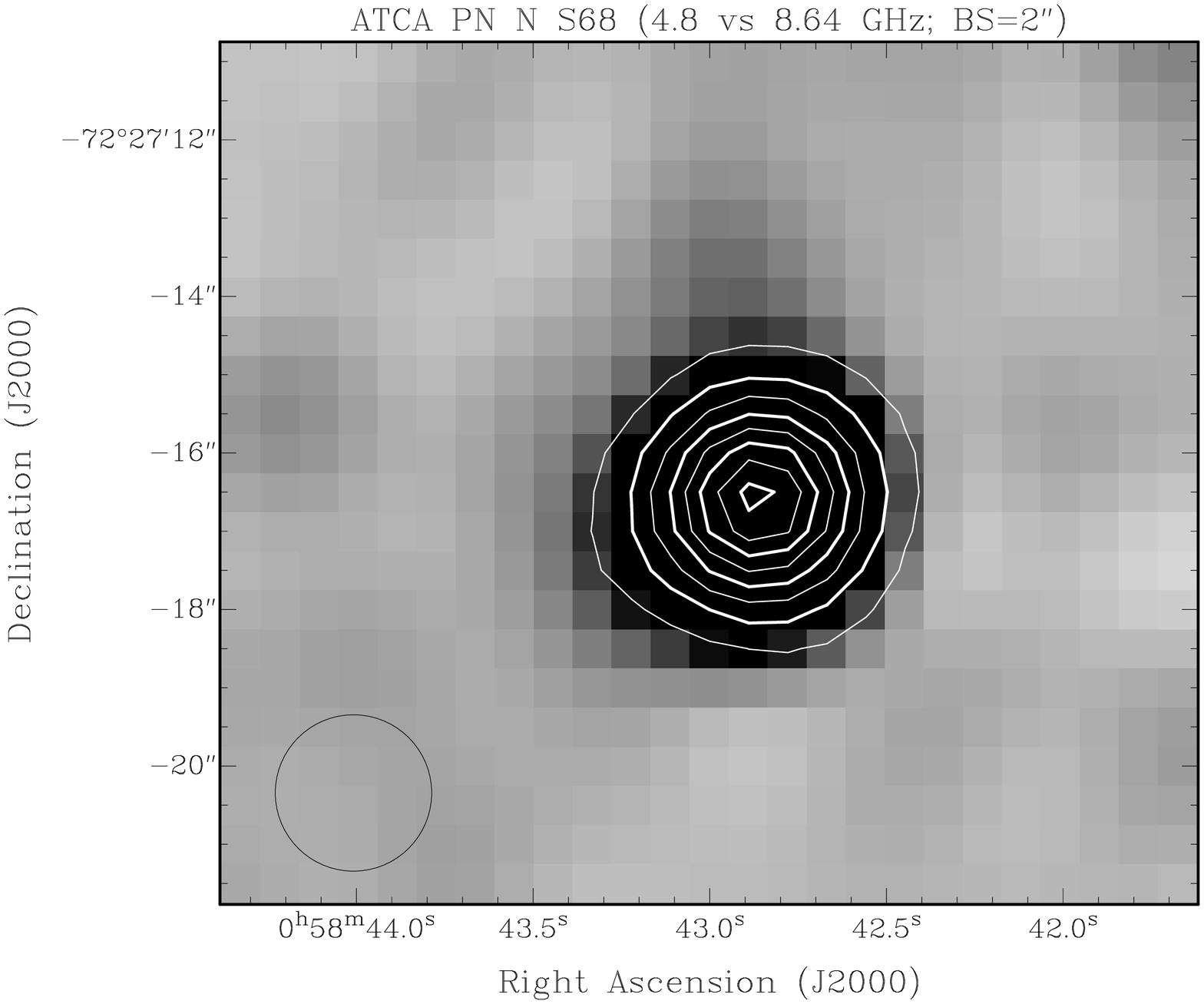}
  \caption{ATCA high resolution radio-continuum images of two SMC PNe -- SMP\,S11 (left) and N\,S68 (right). Gray scale images are at 4.8~GHz and overlaid contours are from 8.64~GHz observations. The radio-continuum contours from 8.64~GHz observations are at both images from 0.5~mJy~beam$^{-1}$ in steps of 0.5~mJy~beam$^{-1}$. All images have rms noise (1~$\sigma$) in order of $\sim$0.1~mJy~beam$^{-1}$. Both synthesised beams are circular (2/4\arcsec) and they are displayed as a black circle in a bottom left corner.}
 \label{fig2-smc}
 \end{center}
\end{figure*}

The radio-continuum surveys described in Section~2 were initially searched within 2\arcsec\ of known optical PNe for co-identifications. In the SMC, PNe lists given by \citet[][his Table~3]{1995A&AS..112..445M} and \citet[][their Table~4]{2002AJ....123..269J}, contain a total of 139 PNe. We found four radio sources (3\%) that were spatially coincident with the PNe: JD~04 (Fig.~\ref{fig1-smc}; left), SMP~S11 (Fig.~\ref{fig2-smc}; left), SMP~S17 (Fig.~\ref{fig1-smc}; right) and N~S68 (Fig.~\ref{fig2-smc}; right). For more details see Table~1. We refer to the PNe using the names listed in \citet{2002AJ....123..269J}. Three other previously classified PNe; MA~1796, MA~1797 \citep{1993A&AS..102..451M} and MG~2 \citep{1995A&AS..112..445M} also appear to be coincident with the corresponding radio sources. However, \citet{2003ApJ...598.1000S} found that these three sources are in fact ultra-compact \HII\ regions and not bona fide PNe. We also note that the radio flux densities for these three objects are much higher than reported here for MC radio PNe. Another previously classified PN in the SMC known as JD\,26 was also detected across the range of our radio frequencies but after close examination we re-classified this object as an \HII\ region.

Within the LMC, we found 11 co-identifications using optical PNe catalogues presented by \citet{1997A&AS..121..407L} and \citet{2006MNRAS.365..401R,2006MNRAS.373..521R}. The catalogue by \citet{1997A&AS..121..407L} contains accurate positions and finding charts for $\sim 280$ LMC PNe compiled from all major surveys prior to 1997. More recent catalogues presented by \citet{2006MNRAS.365..401R,2006MNRAS.373..521R} identify $\sim629$ LMC PNe and PNe candidates, classified into three groups; True (T), Likely (L), and Possible (P). All of our 11 co-identifications are classified type ``T''. We note that there are 11 other radio-continuum sources in the LMC previously classified by \citet{2006MNRAS.365..401R} as PNe or candidates. Namely, two sources were classified as true (RP\,1534 and RP\,105), three as likely (RP\,872, RP\,993 and RP\,1541), and six as possible (RP\,641, RP\,1113, RP\,1495, RP\,1716, RP\,1933 and RP\,2194). \citet{2009IAUS..256...36R} eliminated RP\,1534 and RP\,105 as being true PNe when the MIR data was examined. Eight of these 11 point sources have significantly higher than expected flux densities, in the range of 10-15~mJy, at 1.4~GHz. By applying multi-frequency criteria to these 11 candidates, including the higher than expected flux density, they were found to be compact \HII\ regions. We suggest that an upper limit on radio flux density be included as a new parameter in the multi-frequency PNe selection criteria. More details about this post-facto classification will be given below and in our subsequent papers.

Finally, high resolution imaging and spectra from a recent Hubble Space Telescope (HST) survey of 59 PNe in the MCs \citep{2006ApJS..167..201S}, have 10 (all in the LMC) radio PNe matches (see Table~1, Col.~11), displaying a wide range of morphologies. We note that our 4 PNe radio-continuum detections in the SMC are not yet observed with the HST.

We employed the ``shift technique'' in order to estimate the possibility of false detections. We offset the positions of known optical PNe by 30\arcmin\ in each of four directions ($\pm$RA and $\pm$Dec) and counted the number of spurious identifications. We find only one false detection per Cloud, implying that at most one of the cross-identification in each Cloud that we report here occurred by chance.

Table~1 summarizes our 15 radio detections coincident with known LMC and SMC PNe. We list ATCA radio source name, radio positions (J2000), flux densities at 843~MHz, 1.4~GHz, 2.37~GHz, 4.8~GHz and 8.64~GHz, spectral index (which is defined as $\alpha$ in $S_{\nu} \propto \nu^{\alpha}$, where ($S_{\nu}$) is flux density and ($\nu$) is frequency) and error, flux density ratio between Spitzer MIR at 8$\mu$m and S$_{1.4\mathrm{GHz}}$, optical flux, optical diameter (arcsec) and optical name. None of these sources can be considered radio extended, given our radio resolutions and their expected sizes at the distance of the MCs. All 15 radio detections are within 1\arcsec\ of the best optical positions.

 \subsection{Spitzer Detections and Mid-Infrared properties}

True radio detections of MC PNe are dependent on accurate optical and IR identifications as PNe. For example, \citet{1978PASP...90..621S} note that a faint star with strong H$\alpha$ emission can be misconstrued as a PN if its continuum lies below the sensitivity of the detector. Ultra-compact \HII\ regions have also been confused with PNe and their paper lists examples of both in the LMC and SMC. Some of the optical counterparts among our radio-continuum PN candidates could, therefore, be confused with ultra-compact \HII\ regions as shown by \citet{2003ApJ...598.1000S}.

Compact \HII\ regions and Symbiotics constitute a major contaminant in the search for PNe at all frequencies. However, the mid-infrared morphologies and false colors of \HII\ regions are distinct from those of PNe \citep{2007MNRAS.374..979C,2007ApJ...669..343C}. \HII\ regions that are classified as compact or even ultra-compact are associated with MIR structures such as multiple filaments and/or haloes. Unlike PNe the MIR morphology of \HII\ regions is highly irregular. Their false colors (using IRAC bands {2~(4.5\,$\mu$m), 3~(5.8\,$\mu$m), 4~(8\,$\mu$m)} as blue, green and red, respectively) are generally white, indicative of thermal emission by warm dust grains rather than of fluorescent polycyclic aromatic hydrocarbon bands or molecular hydrogen lines that cause many PNe to appear orange or red.

We have cut out small regions of Spitzer IRAC images around each of the catalogued PNe with potential radio detections. For the LMC, these come from the enhanced products of the SAGE Legacy program (PID~20203) available at the Spitzer Science Center (SSC). For the SMC we downloaded the SSC IRAC mosaics recently available from the SMC SAGE Legacy program (PID~40245). Combining the above techniques with detailed scrutiny of the optical spectra \citep{2006MNRAS.365..401R} we have been able to reclassify the brightest radio detections of PNe candidates (those with flux densities above 10-15~mJy) in both Clouds as coming from \HII\ regions rather than PNe.

\citet{2002AJ....123..269J} list JD~04, SMP~S11, SMP~S17 and N~S68 as previously known SMC PNe in their table~4, recovered by their ``blinking'' technique using images obtained with an \OIII\ filter (on-band) and a nearby continuum filter (off-band). They identified objects as PNe if their diameters were less than 10\arcsec\ and the \OIII\ on-band image flux exceeded twice the off-band flux. We also detect these four objects in our Spitzer images and confirm the status for two (JD~04 and SMP~S17) as bona-fide PNe. The other two objects (SMP~S11 and N~S68), appear to have somewhat larger MIR/RC ratio (see Sect.~4.2) and from the MIR perspective each could also be an ultra-compact \HII\ region or an unusual PN.

\begin{figure}
 \begin{center}
 \includegraphics[width=75mm]{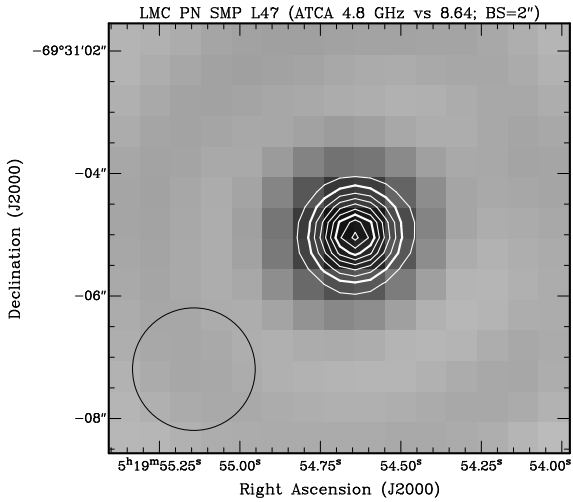}
  \caption{ATCA high resolution radio-continuum image of the LMC PNe -- SMP\,L47. Gray scale image is at 4.8~GHz and overlaid contours are from 8.64~GHz. The radio-continuum contours are from 0.3~mJy~beam$^{-1}$ in steps of 0.2~mJy~beam$^{-1}$. Both images rms noise (1~$\sigma$) are in order of $\sim$0.1~mJy~beam$^{-1}$. The synthesized beam is circular (2\arcsec) and it is displayed as a black circle in a bottom left corner.}
 \label{L47}
 \end{center}
\end{figure}

\begin{figure}
\includegraphics[width=70mm,angle=-90]{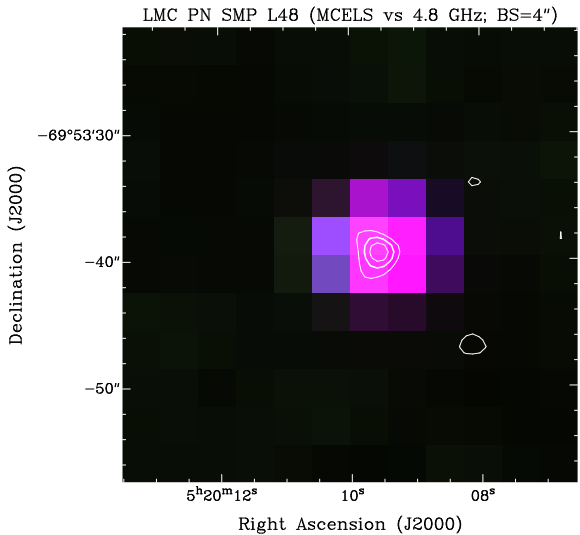}
\caption{The MCELS image of the LMC PNe SMP\,L48 overlaid with the ATCA high resolution radio-continuum image at 4.8~GHz. The radio-continuum contours are from 0.75~mJy~beam$^{-1}$ in steps of 0.25~mJy~beam$^{-1}$. The synthesized beam of the radio image is 4\arcsec\ and the rms noise (1~$\sigma$) is $\sim$0.25~mJy~beam$^{-1}$. We note the large optical extent which is due to the PN faint AGB halos.
  \label{smp48}}
\end{figure}

\begin{figure}
 \begin{center}
 \includegraphics[width=65mm,angle=-90]{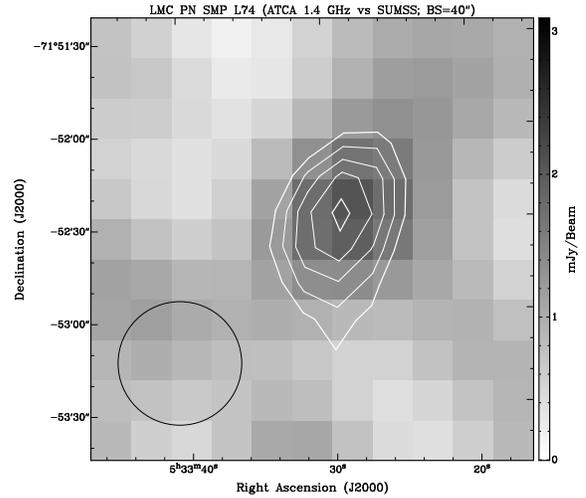}
  \caption{ATCA low resolution radio-continuum image of the LMC PNe -- SMP\,L74. Gray scale image is at 1.4~GHz and overlaid contours are from 843~MHz (SUMSS) observations. The radio-continuum contours are from 2~mJy~beam$^{-1}$ in steps of 0.5~mJy~beam$^{-1}$. The 1.4~GHz image rms noise (1~$\sigma$) is in order of $\sim$0.6~mJy~beam$^{-1}$ and SUMSS $\sim$1~mJy~beam$^{-1}$. Our 1.4~GHz image synthesized beam is circular (40\arcsec) and it is displayed as a black circle in the bottom left corner. }
 \label{fig5-lmc}
 \end{center}
\end{figure}

\begin{figure}
 \begin{center}
 \includegraphics[width=62mm,angle=-90]{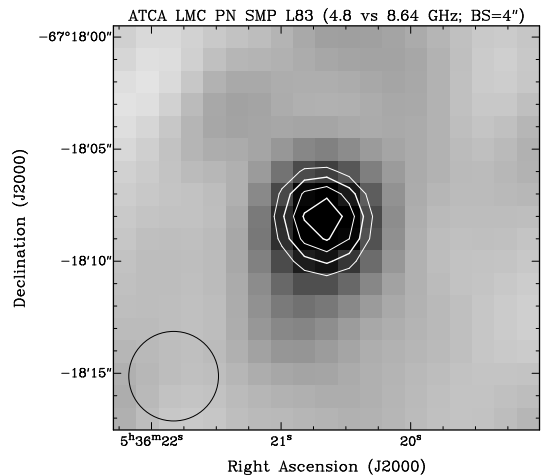}
  \caption{ATCA high resolution radio-continuum image of the LMC PNe -- SMP\,L83. Gray scale image is at 4.8~GHz and overlaid contours (8.64~GHz) are from 0.3~mJy~beam$^{-1}$ in steps of 0.1~mJy~beam$^{-1}$. Both images rms noise (1~$\sigma$) are in order of $\sim$0.1~mJy~beam$^{-1}$. The synthesized beam is circular (4\arcsec) and it is displayed as a black circle in the bottom left corner.}
 \label{L83}
 \end{center}
\end{figure}

SMP\,L8, 25, 33, 39, 47 (Fig.~\ref{L47}), 48 (Fig.~\ref{smp48}), 62, 74 (Fig.~\ref{fig5-lmc}), 83 (Fig.~\ref{L83}), 84 and 89 are originally listed as LMC PNe by \citet{1978PASP...90..621S} based on deep blue and red sensitive objective-prism plates taken from the Curtis Schmidt telescope at the Cerro Tololo Inter-American Observatory. These were originally obtained for other unrelated programs and objects with no evidence of a continuum were selected. Nine of these 11 were confirmed spectroscopically by \citet{2006MNRAS.373..521R}, whilst the other three lie outside their sampled field. \citet{2006ApJS..167..201S} presented comprehensive HST high resolution (spectra and images) of 8 out of 11 LMC radio-continuum PNe. {All 15 radio-continuum PNe detections exhibit canonical optical spectra} leaving us in no doubt that they are indeed optical PNe. {These criteria include the relative sizes and morphologies of nebulae in {\Halpha} and red continuum; the contrast between nebular and ambient {\Halpha} emission; and the presence and intensity ratios in the optical spectrum of forbidden lines, characteristic of PNe but not seen in \HII\ regions. These are the attributes described by \citet{2006MNRAS.373..521R}. More than 10 of our radio detections are also confirmed as PNe from HST imaging 
\citep{2006ApJS..167..201S}.} Also, in the meantime, two other LMC PNe (SMP\,L25 and SMP\,L33) were observed with the HST. {\citet{2008ApJ...672..274B,2009arXiv0905.1124B} have presented two Spitzer spectroscopic studies of a sample of 25 MC PNe from which we report here radio-continuum for 4 objects (SMP\,L8, 62, 83 and SMP\,S11).}

\subsection{Other Radio-continuum Extragalactic PNe and PNe Candidates}

\citet{1994A&A...290..228Z} has reported a radio [WC]-type planetary nebula in the LMC named SMP\,L58 \citep{1978PASP...90..621S} having flux densities of 0.79 and 0.84~mJy at 4.8/8.64~GHz, respectively (based on  {well-calibrated} ATCA June 1993 observations). 
We did not detect SMP\,L58 in any of our LMC mosaic radio surveys as our detection limits (3$\sigma$=1.5~mJy) are well above SMP\,L58 flux densities. These two radio fluxes imply a spectral index of about 0.1$\pm$0.3.

Our non-detection of this PN highlights the difficulty of radio observations of extragalactic PNe, even with the most recent ATCA techniques. 
SMP\,L58 is sufficiently bright among MC PNe that it was originally detected by IRAS at 25~$\mu$m at a level of 220~mJy \citep{1997A&AS..125..419L}, with an estimated error of $\pm$25\%.
The SAGE detection of this PN in the MIPS 24-$\mu$m band corresponds to 190$\pm$5~mJy \citep{2008AJ....135..726H}, consistent with no change in MIR emission over the intervening almost two decades. The 8.0-$\mu$m flux densities in both SAGE epochs (3 months apart) were 37~mJy. That would imply a large MIR/RC ratio of {$\sim$47} if we took a flat spectrum to 1~GHz from \citet{1994A&A...290..228Z} average 4.8/8.65~GHz flux densities or our own measurements. This rather large ratio points to more like \HII\ region nature than a PN. {However, \citet{2009arXiv0905.1124B} present the Spitzer spectrum, and conclude that the dust is carbonaceous, and thus the \HII\ region nature can therefore be
excluded, further highlighting the difficulty of positively identifying PNe even given multi-wavelength data. }

We also note two radio PNe detections in the Sagittarius dwarf galaxy \citep{2000A&A...363..717D}. {When scaled to the distance of the LMC ($\sim$1~mJy at 4.8~GHz) neither of these PNe would be detectable in our surveys.}

\section{Discussion}
 \label{Discussion}
 
 \subsection{Expected Flux Density at the distance of the MCs}

Planetary nebulae within the MCs should not have measurable radio emission much above the sensitivity limits of our present generation data. For example, scaling up the radio fluxes from the very distant (6.6~kpc) Galactic PNe G313.3+00.3 (\citealt{2005ApJ...627..446C}) it would have flux densities of 0.6 and 0.4~mJy at 1.384 and 2.496~GHz at the distance of the LMC. {Another example is the very luminous Galactic PN NGC\,6302. If we adopt expansion distance estimates by \citet{2008MNRAS.385..269M} of $1.17\pm0.14$~kpc and flux density of 3.1~Jy from \citet{2007MNRAS.382.1607C} then this PNe would have a 4.80-GHz flux density at the distance of the LMC of $\sim$1.7~mJy and the SMC of $\sim$1.2~mJy. It is widely accepted \citep[e.g.][]{1990A&A...234..387Z} that the strongest radio PNe in the Galactic Bulge, at a distance of 8.5~kpc, show an upper cutoff flux density of 65~mJy. Scaling this to the distance of the LMC gives an expected flux density of 1.9~mJy and for the SMC 1.3~mJy. 

Probably the most luminous Galactic PNe at the present is NGC\,7027, at a distance of 980$\pm$100~pc \citep{2008ApJ...681.1296Z}, it would have a 4.80-GHz flux density (5.6~Jy at its radio peak in 1987.34) at the distance of the LMC of $\sim$2.2~mJy and the SMC of 1.5~mJy. NGC 7027 is fading, and the original peak flux will have been higher than the current value, although by how much is uncertain. Assuming that a PN central star has a highest likely luminosity of 2$\times$10$^{4}$ L$_{\sun}$ (if higher, the object would evolve too quickly through the PN phase to be detectable), which is about twice that of NGC~7027, suggests a peak LMC radio flux of up to 3~mJy. A further correction factor is needed for the stellar temperature: the radio flux of an optically thick PN varies by up to a factor of 2 during its evolution, caused by the changing stellar temperature (which affects the number of ionizing photons) \citep{2008ApJ...681.1296Z}. According to Zijlstra (2009, priv. com.), this may add another 25-50\% to the peak radio flux. However, very few PNe would be expected to show such values. Almost half of our sample presented here (7 out of 15) have similar or significantly higher flux densities (up to 4.1~mJy of the SMC~PN~N\,S68) at 4.8~GHz than projected values for NGC~7027. This suggest that the most radio luminous PNe such as N\,S68 may represent an upper (or close to upper) limit in radio peak emission. At the present, this SMC PN (N\,S68) is factor of $\sim$3 more luminous than Galactic PN NGC\,7027.}

{In general,} we have found much higher flux densities {then expected} in both MCs PNe; e.g. LMCPN~J054237-700930 (SMP\,L89) with a 1.4-GHz value of 3.1~mJy ($\pm$10\%) and SMCPN~J004336-730227 (JD\,04) with 5.1~mJy at 1.4~GHz. We note that 3 out of 4 detections in the SMC are surprisingly stronger {(by up to a factor of three)} than their LMC and Galactic cousins even though the SMC is some 10~kpc further away then the LMC. While these are small sample statistics one could ask why is the SMC PN sample is brighter than the LMC? Could we be missing even more brighter LMC ones? However, the existence of the observable infrared emission for the four of the five detected SMC PNe imply that the dust in the shell must be relatively dense and close to the central star (CS) in order to be efficiently heated. The major part of the detected radio-continuum emission most likely originates from the dense ionized shell ($<$0.1~pc) which is, at the distance of 60~kpc, much smaller than our 8.64~GHz synthesized beam. The radio spectrum distribution of JD\,04 is fairly flat throughout the observed radio frequency range. However, N\,S68, SMP\,S17 and SMP\,S11 demonstrate a mild but distinctive drop in the flux density toward the lower frequencies which is very likely the effect of the increased optical depth. The spectral index estimate for the N\,S68 in the $\lambda=13$~cm range agrees with the value predicted by the unbounded wind shell model of 0.6 \citep{1975MNRAS.170...41W}. Following the CS evolutionary model of \citet{1981A&A...103..119S} it can be seen that high core mass CS are traveling through the heating part of the HR diagram much faster than the low mass CS. The CS with mass of 0.84~M$_{\sun}$ will pass through the heating stage almost 10 times quicker than the CS with 0.6~M$_{\sun}$ \citep{1993AcA....43..297S}. At the same time, number of ionizing photons ($\lambda<912$\AA) is significantly larger for the high mass CS (for example a 0.65~M$_{\sun}$ central star will produce almost 3 times more ionizing photons than central star with 0.6~M$_{\sun}$ mass). Therefore, assuming that the SMC CS mass distribution is shifted toward 0.65~M$_{\sun}$, it is reasonable to assume that the sudden drop in the radio-continuum flux density and consequent gap between the bright detected objects and the sensitivity limit (see Sec.~4.3) could be caused by the quick recombination phase when the ionizing source is ``turned off''. However, this model will also imply higher densities ($>10^{4}$) in the ionized shell and therefore characteristically optically thick radio-continuum emission at the lower frequencies.

 \subsection{MC PNe Properties and Selection Criteria}

We estimate spectral indices for 11 of the sources shown in Table~1 (Col.~9). Despite the rather large estimated errors, most of our sample is within the expected (--0.3$<\alpha<$+0.3) range. These large errors are most likely due to the low flux density levels of the associated detections. We cannot determine more accurate spectral indices without observations of higher spatial resolution and sensitivity, but we assume that PNe emission is predominantly thermal. However, thermal emission also characterizes compact \HII\ regions. We point out that the radio-continuum flux densities peak at different frequencies and in most cases that the spectral index cannot be described as a straight line. We compared corresponding PNe H$\alpha$ fluxes (Reid \& Parker 2009; in prep.) with radio/IR and found no obvious correlation. Also, our 10 radio LMC PNe detections observed with the HST exhibit a wide range of diameters from very small ($\sim0.08$~pc or 0.32\arcsec; SMP~L47) to very large ($\sim1$~pc or 4\arcsec; SMP~L83; Fig.~\ref{L83}).

While one cannot classify radio sources based solely on spectral index, we note that two of our radio-continuum PNe (SMP\,L74 and SMP\,L83; Figs.~\ref{fig5-lmc} and \ref{L83}) have a very steep spectrum ($\alpha=-0.6\pm0.4$ and $\alpha=-0.5\pm0.2$) {implying} non-thermal emission although the error in the index is large. {We point out that \citet{2004A&A...419..583P} report on large variability in the optical star in SMP\,L83, which may be responsible for the steeper spectral index.} Similarly steep spectral indices may originate from SNRs and background sources such as AGNs and/or quasars. {We exclude these possibilities as they would have very different characteristics at other frequencies such as X-ray and optical. Also, these may exhibit similar physical process as the Galactic PN associated with V1018~Sco and GK~Per with spectral index of --0.8. \citet{2006MNRAS.369..189C} attributed this rather steep spectrum to the collision between the fast and slow winds in this nebula and neither is normally classified as PN. }

Nor is the ratio of MIR to radio flux densities, MIR/RC, uniquely diagnostic for individual objects. The MIR/RC ratios for the MCs are based on MIR flux densities at 8.0$\mu$m  and radio-continuum values at 1.4~GHz. These are given in Table~1 (Col.~10). If there are data, but not at 1.4~GHz, then we assume a flat radio spectrum. The median value for 137 Galactic PNe gives the MIR/RC ratio as 4$\pm$1 {(Cohen et al. 2009, in prep.)}. Very large values, like 50-300, suggest optically thick radio emission regions \citep{2007MNRAS.374..979C}; diffuse \HII\ -- 25 \citep{2007ApJ...669..343C}; ultra-compact \HII\ -- 42 (Murphy et al. in prep). The 14 MCs MIR/RC ratios have a median value of 9$\pm$2; consistent with the Galactic sample (the formal difference between the median values for MC and Galactic PNe has less than 2$\sigma$ significance). An absent ratio in Table~1 indicates no MIR detection.

Although the primary emission mechanism of PN is thermal, \citet{1998ApJ...499L..83D} presented a revised model of the PNe emission mechanism after the discovery of an inner region of non-thermal radio emission in the ``born-again" PNe, A30. They assumed that the fast wind from the central star carries a very weak magnetic field. Interactions of the wind with dense condensations trap magnetic field lines for long periods and stretch them, leading to a strong magnetic field. As the fast wind is shocked, relativistic particles form and interact with the magnetic field to create non-thermal emission. Nonetheless, the flux density from this mechanism for PNe in the MCs would be exceedingly low; less than 1~$\mu$Jy at the distance of the SMC.

\citet{2007ApJ...656..831V} have found that the average central star mass of a sample of 54 PNe located in the LMC is 0.65$\pm$0.07~M$_{\odot}$, slightly higher than reported for those in the Galaxy. They attributed this difference to the lower metallicity in the LMC (on average by half) than in our Galaxy. This naturally raises the question: do MCs PNe evolve differently when compared to their Galactic cousins? 

{\citet{2004MNRAS.348L..23Z} proposed that low-metallicity stars evolve to higher final masses. There has been no convincing confirmation of this (see \citet{2007A&A...467L..29G} for a discussion of the accuracy of mass determinations.) However, our results presented here suggest that this effect is present for the brightest sources. Also, one could interpret this as evidence for a more recent epoch of strong star formation in the MCs, leading to an overabundance of high-mass central star PNe.}  

Our radio PNe detections (Table~1) represent only \mbox{$\sim3$\%} of the optical PNe population of the MCs. Whatever the emission mechanism, we are selecting only the strongest radio-continuum emitters, possibly at a variety of different stages of their evolution \citep{2009arXiv0905.1844V}. A much higher percentage of known Galactic PNe have radio counterparts (for example, from $\sim$1300 PNe from \citet{2003A&A...408.1029K} $\sim$60\% have been detected at 6~cm and $\sim$70\% at 20~cm) (Boji\v ci\'c et al., priv. comm.; \citealt{1998ApJS..117..361C}; \citealt{2005ApJS..159..282L}), presumably because they are much closer.

\subsection{PN Central Star Properties}

Most PNe have central-star and nebular masses of only about 0.6 and 0.3 M$_\odot$, respectively. Detection of white dwarfs in open clusters suggests that the main-sequence mass of PNe progenitors can be as high as 8~M$_\odot$ \citep{1994PASP..106..344K}. Our preliminary spectroscopic study contained 3 of 4 SMC and 10 of 11 LMC radio-detected PNe \citep{2008SerAJ.176...65P,2008SerAJ.177...53P} {and} suggests that nebular electron temperatures are also within the expected range assuming an average density of $10^3$ cm$^{-3}$. Given the values of radio flux density at $\sim5$~GHz, we estimated that the ionized nebular mass of these 13 MCs PNe may be 2.6~$M_\odot$ or greater. {However, forbidden lines are insensitive to high densities and one might question whether the \SII\ densities should be applied to the radio region.}

This study suggests that the MC PNe detected in the radio-continuum may represent a predicted link to the ``missing-mass" problem associated with systems possessing a 1--8 M$_\odot$ central star. If a high rate of mass loss continues for an extended fraction of the Asymptotic Giant Branch (AGB) Star's lifetime, a significant fraction of a star's original mass can be accumulated in a circumstellar envelope (CSE). If the transition from the AGB to PNe stage is short, then such CSEs could have a significant influence on the formation of PNe, resulting in the detection of optical AGB haloes. The presence of these haloes has been known since the 1930s \citep{1937ApJ....86..496D}. 

{We consider the notion that} our ATCA observations may be detecting the extended radio counterparts of these AGB haloes, presumed to be composed of weakly ionized material. {However, the peak radio flux in a PN occurs while the nebula is optically thick for ionizing radiation, meaning only a fraction (in most cases a small fraction) of the circumstellar matter would be ionized. Once the ionized mass increases above ($>$0.1~M$_{\odot}$), the radio flux decreases quite rapidly as the nebulae become optically thin. Therefore, we conclude that haloes are very weak in the radio and cannot be detected in the MCs with the current observations. Although, the ionized mass will increase at a much later stage, when the ISM interaction region merges with the PN \citep{2007MNRAS.382.1233W}, at this time the expected radio flux will be very low indeed.}

While most of the MCs SNRs flux densities are much higher then these 15 PNe \citep[see][for a small sample of MC SNR flux densities]{2007MNRAS.378.1237B,2008SerAJ.176...59C,2008SerAJ.177...61C,2008A&A...485...63F}, they are still bit more luminous then their galactic counterparts ``prompting'' us to call these sources Super PNe or mini-SNRs. While obvious differences remain between the two classes, the physical processes within these two groups (PNe vs SNRs) are perhaps not too dissimilar, if one takes into consideration the fact that older SNRs shock fronts are isothermal in nature.

\section{Conclusion and Future Observations}
  \label{conc}

We present the first 15 extragalactic radio PNe -- all in the MCs. At least 10 of these candidates can currently be positively identified as a PN via high-resolution optical (HST) imaging. All 15 radio-continuum objects examined here, exhibit typical PN characteristics ie. canonical optical spectra and MIR properties leaving us no doubt that they are bone fide. Assuming they are radio PNe, their higher than expected flux densities at lower frequencies are most likely related either to environmental factors, and/or selection effects. We tentatively call this PNe (sub)sample ``Super PNe''.

We are presently conducting high resolution ($\sim1$\arcsec) ATCA observations of all these 15 PNe candidates using a variety of frequencies and arrays. We also plan further optical confirmations of these PNe using high resolution \OIII\ images from the HST. 

\section*{Acknowledgments}

We used the Karma/MIRIAD software packages developed by the ATNF. The Australia Telescope Compact Array is part of the Australia Telescope which is founded by the Commonwealth of Australia for operation as a National Facility managed by the CSIRO. M.C. thanks NASA for funding his participation in this work through ADP grant NNG04GD43G and JPL contract 1320707 with UC Berkeley. We thank the Magellanic Clouds Emission Line Survey (MCELS) team for access to the optical images. {We thank the referee (Albert Zijlstra) for his excellent comments that have greatly improved this manuscript. }

\bibliographystyle{mn2e}
\bibliography{MC_radio_PNs}

\begin{thebibliography}{}

\bibitem[\protect\citeauthoryear{{Beintema}, {van Hoof}, {Lahuis}, {Pottasch},
  {Waters}, {de Graauw}, {Boxhoorn}, {Feuchtgruber} \& {Morris}}{{Beintema}
  et~al.}{1996}]{1996A&A...315L.253B}
{Beintema} D.~A.,  {van Hoof} P.~A.~M.,  {Lahuis} F.,  {Pottasch} S.~R.,
  {Waters} L.~B.~F.~M.,  {de Graauw} T.,  {Boxhoorn} D.~R.,  {Feuchtgruber} H.,
     {Morris} P.~W.,  1996, \aap, 315, L253

\bibitem[\protect\citeauthoryear{{Bernard-Salas}, {Peeters}, {Sloan},
  {Gutenkunst}, {Matsuura}, {Tielens}, {Zijlstra} \& {Houck}}{{Bernard-Salas}
  et~al.}{2009}]{2009arXiv0905.1124B}
{Bernard-Salas} J.,  {Peeters} E.,  {Sloan} G.~C.,  {Gutenkunst} S.,
  {Matsuura} M.,  {Tielens} A.~G.~G.~M.,  {Zijlstra} A.~A.,    {Houck} J.~R.,
  2009, ArXiv e-prints

\bibitem[\protect\citeauthoryear{{Bernard-Salas}, {Pottasch}, {Gutenkunst},
  {Morris} \& {Houck}}{{Bernard-Salas} et~al.}{2008}]{2008ApJ...672..274B}
{Bernard-Salas} J.,  {Pottasch} S.~R.,  {Gutenkunst} S.,  {Morris} P.~W.,
  {Houck} J.~R.,  2008, \apj, 672, 274

\bibitem[\protect\citeauthoryear{{Bock}, {Large} \& {Sadler}}{{Bock}
  et~al.}{1999}]{1999AJ....117.1578B}
{Bock} D.~C.-J.,  {Large} M.~I.,    {Sadler} E.~M.,  1999, \aj, 117, 1578

\bibitem[\protect\citeauthoryear{{Boji{\v c}i{\'c}}, {Filipovi{\'c}}, {Parker},
  {Payne}, {Jones}, {Reid}, {Kawamura} \& {Fukui}}{{Boji{\v c}i{\'c}}
  et~al.}{2007}]{2007MNRAS.378.1237B}
{Boji{\v c}i{\'c}} I.~S.,  {Filipovi{\'c}} M.~D.,  {Parker} Q.~A.,  {Payne}
  J.~L.,  {Jones} P.~A.,  {Reid} W.,  {Kawamura} A.,    {Fukui} Y.,  2007,
  \mnras, 378, 1237

\bibitem[\protect\citeauthoryear{{Casassus}, {Nyman}, {Dickinson} \&
  {Pearson}}{{Casassus} et~al.}{2007}]{2007MNRAS.382.1607C}
{Casassus} S.,  {Nyman} L.-{\AA}.,  {Dickinson} C.,    {Pearson} T.~J.,  2007,
  \mnras, 382, 1607

\bibitem[\protect\citeauthoryear{{Cohen}, {Chapman}, {Deacon}, {Sault},
  {Parker} \& {Green}}{{Cohen} et~al.}{2006}]{2006MNRAS.369..189C}
{Cohen} M.,  {Chapman} J.~M.,  {Deacon} R.~M.,  {Sault} R.~J.,  {Parker} Q.~A.,
     {Green} A.~J.,  2006, \mnras, 369, 189

\bibitem[\protect\citeauthoryear{{Cohen}, {Green}, {Meade}, {Babler},
  {Indebetouw}, {Whitney}, {Watson}, {Wolfire}, {Wolff}, {Mathis} \&
  {Churchwell}}{{Cohen} et~al.}{2007a}]{2007MNRAS.374..979C}
{Cohen} M.,  {Green} A.~J.,  {Meade} M.~R.,  {Babler} B.,  {Indebetouw} R.,
  {Whitney} B.~A.,  {Watson} C.,  {Wolfire} M.,  {Wolff} M.~J.,  {Mathis}
  J.~S.,    {Churchwell} E.~B.,  2007a, \mnras, 374, 979

\bibitem[\protect\citeauthoryear{{Cohen}, {Green}, {Roberts}, {Meade},
  {Babler}, {Indebetouw}, {Whitney}, {Watson}, {Wolfire}, {Wolff}, {Mathis} \&
  {Churchwell}}{{Cohen} et~al.}{2005}]{2005ApJ...627..446C}
{Cohen} M.,  {Green} A.~J.,  {Roberts} M.~S.~E.,  {Meade} M.~R.,  {Babler} B.,
  {Indebetouw} R.,  {Whitney} B.~A.,  {Watson} C.,  {Wolfire} M.,  {Wolff}
  M.~J.,  {Mathis} J.~S.,    {Churchwell} E.~B.,  2005, \apj, 627, 446

\bibitem[\protect\citeauthoryear{{Cohen}, {Parker}, {Green}, {Murphy},
  {Miszalski}, {Frew}, {Meade}, {Babler}, {Indebetouw}, {Whitney}, {Watson},
  {Churchwell} \& {Watson}}{{Cohen} et~al.}{2007b}]{2007ApJ...669..343C}
{Cohen} M.,  {Parker} Q.~A.,  {Green} A.~J.,  {Murphy} T.,  {Miszalski} B.,
  {Frew} D.~J.,  {Meade} M.~R.,  {Babler} B.,  {Indebetouw} R.,  {Whitney}
  B.~A.,  {Watson} C.,  {Churchwell} E.~B.,    {Watson} D.~F.,  2007b, \apj,
  669, 343

\bibitem[\protect\citeauthoryear{{Condon} \& {Kaplan}}{{Condon} \&
  {Kaplan}}{1998}]{1998ApJS..117..361C}
{Condon} J.~J.,  {Kaplan} D.~L.,  1998, \apjs, 117, 361

\bibitem[\protect\citeauthoryear{{Crawford}, {Filipovic}, {de Horta},
  {Stootman} \& {Payne}}{{Crawford} et~al.}{2008b}]{2008SerAJ.177...61C}
{Crawford} E.~J.,  {Filipovic} M.~D.,  {de Horta} A.~Y.,  {Stootman} F.~H.,
  {Payne} J.~L.,  2008b, Serbian Astronomical Journal, 177, 61

\bibitem[\protect\citeauthoryear{{Crawford}, {Filipovic} \& {Payne}}{{Crawford}
  et~al.}{2008a}]{2008SerAJ.176...59C}
{Crawford} E.~J.,  {Filipovic} M.~D.,    {Payne} J.~L.,  2008a, Serbian
  Astronomical Journal, 176, 59

\bibitem[\protect\citeauthoryear{{Dgani} \& {Soker}}{{Dgani} \&
  {Soker}}{1998}]{1998ApJ...499L..83D}
{Dgani} R.,  {Soker} N.,  1998, \apjl, 499, L83+

\bibitem[\protect\citeauthoryear{{Dickel}, {Gruendl}, {McIntyre}, {Amy} \&
  {Milne}}{{Dickel} et~al.}{2009}]{2009IAUS..256...14D}
{Dickel} J.,  {Gruendl} R.~A.,  {McIntyre} V.,  {Amy} S.,    {Milne} D.,  2009,
  in {van Loon} J.~T.,  {Oliveira} J.~M.,  eds, IAU Symposium Vol.~256 of IAU
  Symposium, {Survey of the Magellanic Clouds at 4.8 and 8.64 GHz}.
pp 14--19

\bibitem[\protect\citeauthoryear{{Dickel}, {McIntyre}, {Gruendl} \&
  {Milne}}{{Dickel} et~al.}{2005}]{2005AJ....129..790D}
{Dickel} J.~R.,  {McIntyre} V.~J.,  {Gruendl} R.~A.,    {Milne} D.~K.,  2005,
  \aj, 129, 790

\bibitem[\protect\citeauthoryear{{Dudziak}, {P{\'e}quignot}, {Zijlstra} \&
  {Walsh}}{{Dudziak} et~al.}{2000}]{2000A&A...363..717D}
{Dudziak} G.,  {P{\'e}quignot} D.,  {Zijlstra} A.~A.,    {Walsh} J.~R.,  2000,
  \aap, 363, 717

\bibitem[\protect\citeauthoryear{{Duncan}}{{Duncan}}{1937}]{1937ApJ....86..496%
D}
{Duncan} J.~C.,  1937, \apj, 86, 496

\bibitem[\protect\citeauthoryear{{Filipovi{\'c}}, {Bohlsen}, {Reid},
  {Staveley-Smith}, {Jones}, {Nohejl} \& {Goldstein}}{{Filipovi{\'c}}
  et~al.}{2002}]{2002MNRAS.335.1085F}
{Filipovi{\'c}} M.~D.,  {Bohlsen} T.,  {Reid} W.,  {Staveley-Smith} L.,
  {Jones} P.~A.,  {Nohejl} K.,    {Goldstein} G.,  2002, \mnras, 335, 1085

\bibitem[\protect\citeauthoryear{{Filipovic}, {Crawford}, {Hughes}, {Leverenz},
  {de Horta}, {Payne}, {Staveley-Smith}, {Dickel}, {Stootman} \&
  {White}}{{Filipovic} et~al.}{2009}]{2009IAUS..256...PDF-8}
{Filipovic} M.~D.,  {Crawford} E.~J.,  {Hughes} A.,  {Leverenz} H.,  {de Horta}
  A.~Y.,  {Payne} J.~L.,  {Staveley-Smith} L.,  {Dickel} J.~R.,  {Stootman}
  F.~H.,    {White} G.~L.,  2009, in {van Loon} J.~T.,  {Oliveira} J.~M.,  eds,
  IAU Symposium Vol.~256 of IAU Symposium, {New radio-continuum surveys of the
  Magellanic Clouds}.
pp PDF--8

\bibitem[\protect\citeauthoryear{{Filipovi{\'c}}, {Haberl}, {Winkler},
  {Pietsch}, {Payne}, {Crawford}, {de Horta}, {Stootman} \&
  {Reaser}}{{Filipovi{\'c}} et~al.}{2008}]{2008A&A...485...63F}
{Filipovi{\'c}} M.~D.,  {Haberl} F.,  {Winkler} P.~F.,  {Pietsch} W.,  {Payne}
  J.~L.,  {Crawford} E.~J.,  {de Horta} A.~Y.,  {Stootman} F.~H.,    {Reaser}
  B.~E.,  2008, \aap, 485, 63

\bibitem[\protect\citeauthoryear{{Filipovic}, {Haynes}, {White}, {Jones},
  {Klein} \& {Wielebinski}}{{Filipovic} et~al.}{1995}]{1995A&AS..111..311F}
{Filipovic} M.~D.,  {Haynes} R.~F.,  {White} G.~L.,  {Jones} P.~A.,  {Klein}
  U.,    {Wielebinski} R.,  1995, \aaps, 111, 311

\bibitem[\protect\citeauthoryear{{Filipovic}, {Jones}, {White}, {Haynes},
  {Klein} \& {Wielebinski}}{{Filipovic} et~al.}{1997}]{1997A&AS..121..321F}
{Filipovic} M.~D.,  {Jones} P.~A.,  {White} G.~L.,  {Haynes} R.~F.,  {Klein}
  U.,    {Wielebinski} R.,  1997, \aaps, 121, 321

\bibitem[\protect\citeauthoryear{{Filipovi{\'c}}, {Payne}, {Reid}, {Danforth},
  {Staveley-Smith}, {Jones} \& {White}}{{Filipovi{\'c}}
  et~al.}{2005}]{2005MNRAS.364..217F}
{Filipovi{\'c}} M.~D.,  {Payne} J.~L.,  {Reid} W.,  {Danforth} C.~W.,
  {Staveley-Smith} L.,  {Jones} P.~A.,    {White} G.~L.,  2005, \mnras, 364,
  217

\bibitem[\protect\citeauthoryear{{Gesicki} \& {Zijlstra}}{{Gesicki} \&
  {Zijlstra}}{2007}]{2007A&A...467L..29G}
{Gesicki} K.,  {Zijlstra} A.~A.,  2007, \aap, 467, L29

\bibitem[\protect\citeauthoryear{{Herrmann}, {Ciardullo}, {Feldmeier} \&
  {Vinciguerra}}{{Herrmann} et~al.}{2008}]{2008ApJ...683..630H}
{Herrmann} K.~A.,  {Ciardullo} R.,  {Feldmeier} J.~J.,    {Vinciguerra} M.,
  2008, \apj, 683, 630

\bibitem[\protect\citeauthoryear{{Hora}, {Cohen}, {Ellis}, {Meixner}, {Blum},
  {Latter}, {Whitney}, {Meade}, {Babler}, {Indebetouw}, {Gordon},
  {Engelbracht}, {For}, {Block}, {Misselt}, {Vijh} \& {Leitherer}}{{Hora}
  et~al.}{2008}]{2008AJ....135..726H}
{Hora} J.~L.,  {Cohen} M.,  {Ellis} R.~G.,  {Meixner} M.,  {Blum} R.~D.,
  {Latter} W.~B.,  {Whitney} B.~A.,  {Meade} M.~R.,  {Babler} B.~L.,
  {Indebetouw} R.,  {Gordon} K.,  {Engelbracht} C.~W.,  {For} B.-Q.,  {Block}
  M.,  {Misselt} K.,  {Vijh} U.,    {Leitherer} C.,  2008, \aj, 135, 726

\bibitem[\protect\citeauthoryear{{Hughes}, {Staveley-Smith}, {Kim}, {Wolleben}
  \& {Filipovi{\'c}}}{{Hughes} et~al.}{2007}]{2007MNRAS.382..543H}
{Hughes} A.,  {Staveley-Smith} L.,  {Kim} S.,  {Wolleben} M.,
  {Filipovi{\'c}} M.,  2007, \mnras, 382, 543

\bibitem[\protect\citeauthoryear{{Hughes}, {Wong}, {Ekers}, {Staveley-Smith},
  {Filipovic}, {Maddison}, {Fukui} \& {Mizuno}}{{Hughes}
  et~al.}{2006}]{2006MNRAS.370..363H}
{Hughes} A.,  {Wong} T.,  {Ekers} R.,  {Staveley-Smith} L.,  {Filipovic} M.,
  {Maddison} S.,  {Fukui} Y.,    {Mizuno} N.,  2006, \mnras, 370, 363

\bibitem[\protect\citeauthoryear{{Jacoby} \& {De Marco}}{{Jacoby} \& {De
  Marco}}{2002}]{2002AJ....123..269J}
{Jacoby} G.~H.,  {De Marco} O.,  2002, \aj, 123, 269

\bibitem[\protect\citeauthoryear{{Kerber}, {Mignani}, {Guglielmetti} \&
  {Wicenec}}{{Kerber} et~al.}{2003}]{2003A&A...408.1029K}
{Kerber} F.,  {Mignani} R.~P.,  {Guglielmetti} F.,    {Wicenec} A.,  2003,
  \aap, 408, 1029

\bibitem[\protect\citeauthoryear{{Kwok}}{{Kwok}}{1994}]{1994PASP..106..344K}
{Kwok} S.,  1994, \pasp, 106, 344

\bibitem[\protect\citeauthoryear{{Kwok}}{{Kwok}}{2000}]{2000oepn.book.....K}
{Kwok} S.,  2000, {The Origin and Evolution of Planetary Nebulae}.
The origin and evolution of planetary nebulae / Sun Kwok.~Cambridge ; New York
  : Cambridge University Press, 2000.~(Cambridge astrophysics series ; 33)

\bibitem[\protect\citeauthoryear{{Kwok}}{{Kwok}}{2005}]{2005JKAS...38..271K}
{Kwok} S.,  2005, Journal of Korean Astronomical Society, 38, 271

\bibitem[\protect\citeauthoryear{{Leisy}, {Dennefeld}, {Alard} \&
  {Guibert}}{{Leisy} et~al.}{1997}]{1997A&AS..121..407L}
{Leisy} P.,  {Dennefeld} M.,  {Alard} C.,    {Guibert} J.,  1997, \aaps, 121,
  407

\bibitem[\protect\citeauthoryear{{Loup}, {Zijlstra}, {Waters} \&
  {Groenewegen}}{{Loup} et~al.}{1997}]{1997A&AS..125..419L}
{Loup} C.,  {Zijlstra} A.~A.,  {Waters} L.~B.~F.~M.,    {Groenewegen} M.~A.~T.,
   1997, \aaps, 125, 419

\bibitem[\protect\citeauthoryear{{Luo}, {Condon} \& {Yin}}{{Luo}
  et~al.}{2005}]{2005ApJS..159..282L}
{Luo} S.~G.,  {Condon} J.~J.,    {Yin} Q.~F.,  2005, \apjs, 159, 282

\bibitem[\protect\citeauthoryear{{Mao}, {Gaensler}, {Stanimirovi{\'c}},
  {Haverkorn}, {McClure-Griffiths}, {Staveley-Smith} \& {Dickey}}{{Mao}
  et~al.}{2008}]{2008ApJ...688.1029M}
{Mao} S.~A.,  {Gaensler} B.~M.,  {Stanimirovi{\'c}} S.,  {Haverkorn} M.,
  {McClure-Griffiths} N.~M.,  {Staveley-Smith} L.,    {Dickey} J.~M.,  2008,
  \apj, 688, 1029

\bibitem[\protect\citeauthoryear{{Meaburn}, {Lloyd}, {Vaytet} \&
  {L{\'o}pez}}{{Meaburn} et~al.}{2008}]{2008MNRAS.385..269M}
{Meaburn} J.,  {Lloyd} M.,  {Vaytet} N.~M.~H.,    {L{\'o}pez} J.~A.,  2008,
  \mnras, 385, 269

\bibitem[\protect\citeauthoryear{{Meyssonnier} \& {Azzopardi}}{{Meyssonnier} \&
  {Azzopardi}}{1993}]{1993A&AS..102..451M}
{Meyssonnier} N.,  {Azzopardi} M.,  1993, \aaps, 102, 451

\bibitem[\protect\citeauthoryear{{Mezger} \& {Henderson}}{{Mezger} \&
  {Henderson}}{1967}]{1967ApJ...147..471M}
{Mezger} P.~G.,  {Henderson} A.~P.,  1967, \apj, 147, 471

\bibitem[\protect\citeauthoryear{{Morgan}}{{Morgan}}{1995}]{1995A&AS..112..445%
M}
{Morgan} D.~H.,  1995, \aaps, 112, 445

\bibitem[\protect\citeauthoryear{{Payne}, {Filipovic}, {Crawford}, {de Horta},
  {White} \& {Stootman}}{{Payne} et~al.}{2008a}]{2008SerAJ.176...65P}
{Payne} J.~L.,  {Filipovic} M.~D.,  {Crawford} E.~J.,  {de Horta} A.~Y.,
  {White} G.~L.,    {Stootman} F.~H.,  2008a, Serbian Astronomical Journal, 176,
  65

\bibitem[\protect\citeauthoryear{{Payne}, {Filipovic}, {Millar}, {Crawford},
  {de Horta}, {Stootman} \& {Urosevic}}{{Payne}
  et~al.}{2008b}]{2008SerAJ.177...53P}
{Payne} J.~L.,  {Filipovic} M.~D.,  {Millar} W.~C.,  {Crawford} E.~J.,  {de
  Horta} A.~Y.,  {Stootman} F.~H.,    {Urosevic} D.,  2008b, Serbian
  Astronomical Journal, 177, 53

\bibitem[\protect\citeauthoryear{{Payne}, {Filipovi{\'c}}, {Reid}, {Jones},
  {Staveley-Smith} \& {White}}{{Payne} et~al.}{2004}]{2004MNRAS.355...44P}
{Payne} J.~L.,  {Filipovi{\'c}} M.~D.,  {Reid} W.,  {Jones} P.~A.,
  {Staveley-Smith} L.,    {White} G.~L.,  2004, \mnras, 355, 44

\bibitem[\protect\citeauthoryear{{Payne}, {Tauber}, {Filipovic}, {Crawford} \&
  {de Horta}}{{Payne} et~al.}{2009}]{2009SerAJ.178...65P}
{Payne} J.~L.,  {Tauber} L.~A.,  {Filipovic} M.~D.,  {Crawford} E.~J.,    {de
  Horta} 2009, Serbian Astronomical Journal, 178, 65

\bibitem[\protect\citeauthoryear{{Pe{\~n}a}, {Hamann}, {Ruiz}, {Peimbert} \&
  {Peimbert}}{{Pe{\~n}a} et~al.}{2004}]{2004A&A...419..583P}
{Pe{\~n}a} M.,  {Hamann} W.-R.,  {Ruiz} M.~T.,  {Peimbert} A.,    {Peimbert}
  M.,  2004, \aap, 419, 583

\bibitem[\protect\citeauthoryear{{Reid} \& {Parker}}{{Reid} \&
  {Parker}}{2006a}]{2006MNRAS.365..401R}
{Reid} W.~A.,  {Parker} Q.~A.,  2006a, \mnras, 365, 401

\bibitem[\protect\citeauthoryear{{Reid} \& {Parker}}{{Reid} \&
  {Parker}}{2006b}]{2006MNRAS.373..521R}
{Reid} W.~A.,  {Parker} Q.~A.,  2006b, \mnras, 373, 521

\bibitem[\protect\citeauthoryear{{Reid} \& {Parker}}{{Reid} \&
  {Parker}}{2009}]{2009IAUS..256...36R}
{Reid} W.~A.,  {Parker} Q.~A.,  2009, in {van Loon} J.~T.,  {Oliveira} J.~M.,
  eds, IAU Symposium Vol.~256 of IAU Symposium, {Significant new planetary
  nebula discoveries as powerful probes of the LMC}.
pp 36--42

\bibitem[\protect\citeauthoryear{{Sanduleak}, {MacConnell} \&
  {Philip}}{{Sanduleak} et~al.}{1978}]{1978PASP...90..621S}
{Sanduleak} N.,  {MacConnell} D.~J.,    {Philip} A.~G.~D.,  1978, \pasp, 90,
  621

\bibitem[\protect\citeauthoryear{{Schoenberner}}{{Schoenberner}}{1981}]{1981A&%
A...103..119S}
{Schoenberner} D.,  1981, \aap, 103, 119

\bibitem[\protect\citeauthoryear{{Schoenberner}}{{Schoenberner}}{1993}]{1993Ac%
A....43..297S}
{Schoenberner} D.,  1993, Acta Astronomica, 43, 297

\bibitem[\protect\citeauthoryear{{Shaw}, {Stanghellini}, {Villaver} \&
  {Mutchler}}{{Shaw} et~al.}{2006}]{2006ApJS..167..201S}
{Shaw} R.~A.,  {Stanghellini} L.,  {Villaver} E.,    {Mutchler} M.,  2006,
  \apjs, 167, 201

\bibitem[\protect\citeauthoryear{{Stanghellini}, {Villaver}, {Shaw} \&
  {Mutchler}}{{Stanghellini} et~al.}{2003}]{2003ApJ...598.1000S}
{Stanghellini} L.,  {Villaver} E.,  {Shaw} R.~A.,    {Mutchler} M.,  2003,
  \apj, 598, 1000

\bibitem[\protect\citeauthoryear{{Turtle}, {Ye}, {Amy} \& {Nicholls}}{{Turtle}
  et~al.}{1998}]{1998PASA...15..280T}
{Turtle} A.~J.,  {Ye} T.,  {Amy} S.~W.,    {Nicholls} J.,  1998, Publications
  of the Astronomical Society of Australia, 15, 280

\bibitem[\protect\citeauthoryear{{Villaver}, {Stanghellini} \&
  {Shaw}}{{Villaver} et~al.}{2007}]{2007ApJ...656..831V}
{Villaver} E.,  {Stanghellini} L.,    {Shaw} R.~A.,  2007, \apj, 656, 831

\bibitem[\protect\citeauthoryear{{Vukotic}, {Urosevic}, {Filipovic} \&
  {Payne}}{{Vukotic} et~al.}{2009}]{2009arXiv0905.1844V}
{Vukotic} B.,  {Urosevic} D.,  {Filipovic} M.~D.,    {Payne} J.~L.,  2009,
  ArXiv e-prints

\bibitem[\protect\citeauthoryear{{Wareing}, {Zijlstra} \& {O'Brien}}{{Wareing}
  et~al.}{2007}]{2007MNRAS.382.1233W}
{Wareing} C.~J.,  {Zijlstra} A.~A.,    {O'Brien} T.~J.,  2007, \mnras, 382,
  1233

\bibitem[\protect\citeauthoryear{{Wright} \& {Barlow}}{{Wright} \&
  {Barlow}}{1975}]{1975MNRAS.170...41W}
{Wright} A.~E.,  {Barlow} M.~J.,  1975, \mnras, 170, 41

\bibitem[\protect\citeauthoryear{{Zijlstra}}{{Zijlstra}}{1990}]{1990A&A...234.%
.387Z}
{Zijlstra} A.~A.,  1990, \aap, 234, 387

\bibitem[\protect\citeauthoryear{{Zijlstra}}{{Zijlstra}}{2004}]{2004MNRAS.348L%
..23Z}
{Zijlstra} A.~A.,  2004, \mnras, 348, L23

\bibitem[\protect\citeauthoryear{{Zijlstra}, {van Hoof}, {Chapman} \&
  {Loup}}{{Zijlstra} et~al.}{1994}]{1994A&A...290..228Z}
{Zijlstra} A.~A.,  {van Hoof} P.~A.~M.,  {Chapman} J.~M.,    {Loup} C.,  1994,
  \aap, 290, 228

\bibitem[\protect\citeauthoryear{{Zijlstra}, {van Hoof} \& {Perley}}{{Zijlstra}
  et~al.}{2008}]{2008ApJ...681.1296Z}
{Zijlstra} A.~A.,  {van Hoof} P.~A.~M.,    {Perley} R.~A.,  2008, \apj, 681,
  1296

\end{thebibliography}
\appendix
\clearpage
\begin{table*}
\vbox to220mm{\vfil Landscape Table to go here
\caption{}
\vfil}
\label{Table1}
\label{lastpage}
\end{table*}
\end{document}